\documentclass[a4paper, 12pt]{article}

\usepackage{tikz}
\usepackage{latexsym,amsmath,amsfonts,amssymb}
\usepackage{amsthm}
\usepackage{mathrsfs}
\usepackage[latin1]{inputenc}
\usepackage[american]{babel}
\usepackage{bbm}
\usepackage{cite}    
\usepackage[pdfencoding=auto]{hyperref}
\hypersetup{colorlinks, citecolor=[rgb]{.7,0,0}, linkcolor=[rgb]{0,0,0.7}, urlcolor=[rgb]{0,0,0.5}}

\newcommand{\parfrac}[2]{\frac{\partial #1}{\partial #2}}

\newcommand{\rep}[1]{\ensuremath{\mathbf{#1}}}
\newcommand{\wt}{\widetilde}

\newcommand{\matht}[1]{\texorpdfstring{\ensuremath{\boldsymbol{#1}}}{#1}}

\newcommand{\ie}{\textit{i.e.}}

\numberwithin{equation}{section}

\newcommand{\nn}{\nonumber}

\newcommand{\be}{\begin{equation}} \newcommand{\ee}{\end{equation}}
\newcommand{\bea}{\begin{equation} \begin{aligned}} \newcommand{\eea}{\end{aligned} \end{equation}}
\newcommand{\ba}{\begin{array}} \newcommand{\ea}{\end{array}}

\newcommand{\cI}{\mathcal{I}}

\newcommand{\cN}{\mathcal{N}}
\newcommand{\cO}{\mathcal{O}}

\newcommand{\cQ}{\mathcal{Q}}

\newcommand{\cZ}{\mathcal{Z}}

\newcommand{\bC}{\mathbb{C}}

\newcommand{\bQ}{\mathbb{Q}}
\newcommand{\bR}{\mathbb{R}}

\newcommand{\bT}{\mathbb{T}}

\newcommand{\bZ}{\mathbb{Z}}

\newcommand{\fq}{\mathfrak{q}}

\newcommand{\fsu}{\mathfrak{su}}

\DeclareMathOperator{\Tr}{Tr}

\DeclareMathOperator{\im}{\mathbb{I}m}

\DeclareMathOperator{\sn}{sn}
\DeclareMathOperator{\cn}{cn}
\DeclareMathOperator{\dn}{dn}
\DeclareMathOperator{\arcsn}{arcsn}

\makeatletter
\def\blfootnote{\gdef\@thefnmark{}\@footnotetext}
\makeatother

\begin{document}

\thispagestyle{empty}
\begin{flushright}
SISSA  07/2021/FISI
\end{flushright}
\vspace{13mm}  
\begin{center}
{\huge  Superconformal index of low-rank gauge \\[.5em] theories via the Bethe Ansatz}
\\[13mm]
{\large Francesco Benini$^{1,2,3}$ and Giovanni Rizi$^{1}$}

\bigskip
{\it
$^1$ SISSA, Via Bonomea 265, 34136 Trieste, Italy \\[.2em]
$^2$ INFN, Sezione di Trieste, Via Valerio 2, 34127 Trieste, Italy \\[.2em]
$^3$ ICTP, Strada Costiera 11, 34151 Trieste, Italy \\[.2em]
}


\vspace{2cm}

{\parbox{16cm}{\hspace{5mm}
We study the Bethe Ansatz formula for the superconformal index, in the case of 4d $\cN=4$ super-Yang-Mills with gauge group $SU(N)$. We observe that not all solutions to the Bethe Ansatz Equations (BAEs) contribute to the index, and thus formulate ``reduced BAEs'' such that all and only their solutions contribute. We then propose, sharpening a conjecture of Arabi Ardehali \textit{et al.} \cite{ArabiArdehali:2019orz}, that there is a one-to-one correspondence between branches of solutions to the reduced BAEs and vacua of the 4d $\cN=1^*$ theory. We test the proposal in the case of $SU(2)$ and $SU(3)$. In the case of $SU(3)$, we confirm that there is a continuous family of solutions, whose contribution to the index is non-vanishing.
}}
\end{center}

\newpage
\pagenumbering{arabic}
\setcounter{page}{1}
\setcounter{footnote}{0}
\renewcommand{\thefootnote}{\arabic{footnote}}



\section{Introduction and results}
\label{sec: intro}

The AdS/CFT correspondence \cite{Maldacena:1997re, Witten:1998qj, Gubser:1998bc} provides us with a non-perturbative definition of quantum gravity in anti-de-Sitter (AdS) space, in terms of an ordinary quantum field theory (QFT) living on its conformal boundary. In particular, ensembles of states in the boundary theory capture the physics of black holes in the bulk \cite{Witten:1998zw}. This means that a counting of quantum microstates in the boundary theory, performed by studying suitable partition functions in the large $N$ limit, can reproduce and explain the Bekenstein-Hawking entropy of black holes in AdS \cite{Benini:2015eyy}. The last few years have seen a lot of activity in this direction, in the context of supersymmetric black holes in which explicit and precise computations can be performed, see for instance \cite{Hosseini:2016tor, Benini:2016hjo, Benini:2016rke, Hosseini:2016cyf, Cabo-Bizet:2017jsl, Azzurli:2017kxo, Hosseini:2017fjo, Benini:2017oxt, Hosseini:2018uzp, Crichigno:2018adf, Suh:2018tul, Hosseini:2018usu, Suh:2018szn, Cabo-Bizet:2018ehj, Choi:2018hmj, Benini:2018ywd, Honda:2019cio, Fluder:2019szh, ArabiArdehali:2019tdm, Kim:2019yrz, Cabo-Bizet:2019osg, Amariti:2019mgp, Gang:2019uay, Kantor:2019lfo, Lezcano:2019pae, Lanir:2019abx, Choi:2019zpz, Bobev:2019zmz, Nian:2019pxj, Cabo-Bizet:2019eaf, Benini:2019dyp, Cabo-Bizet:2020nkr, Murthy:2020rbd, Agarwal:2020zwm, Benini:2020gjh, GonzalezLezcano:2020yeb, Copetti:2020dil, Goldstein:2020yvj, Cabo-Bizet:2020ewf, Amariti:2020jyx}.

An interesting case is that of type IIB string theory (and its supergravity low-energy limit) in AdS$_5 \times S^5$, whose physics is captured by the four-dimensional $\cN=4$ super-Yang-Mills (SYM) theory with gauge group $SU(N)$ \cite{Maldacena:1997re}. The microstates of 1/16 BPS black holes in AdS$_5$ are captured by the superconformal index \cite{Romelsberger:2005eg, Kinney:2005ej}, a supersymmetric partition function that can be computed exactly. The large $N$ limit can be studied with a variety of methods \cite{Sundborg:1999ue, Aharony:2003sx, Kinney:2005ej, Choi:2018hmj, Benini:2018ywd, Cabo-Bizet:2019eaf, Copetti:2020dil}. From it, one can extract the Bekenstein-Hawking entropy, but also find hints of new physics. For instance, one can compute perturbative corrections expected to come from higher-derivative terms \cite{GonzalezLezcano:2020yeb}, as well as non-perturbative corrections coming from Euclidean complex saddles of the gravitational action \cite{BeniniToappear, AharonyTalk20-11} (thus highlighting the importance of Euclidean complex saddles \cite{Cabo-Bizet:2018ehj}) and corrections expected from D-branes. 

A particularly effective way to compute the large $N$ limit is to use the Bethe Ansatz (BA) approach \cite{Benini:2018ywd}, based on the Bethe Ansatz formulation of the superconformal index \cite{Closset:2017bse, Benini:2018mlo}. The standard integral formula for the index \cite{Sundborg:1999ue, Aharony:2003sx, Romelsberger:2005eg, Kinney:2005ej} can be recast as a sum over the solution set to certain transcendental equations, dubbed Bethe Ansatz Equations (BAEs) because of their similarity with the ones appearing in the context of integrable systems. One particular solution to the BAEs reproduces the Bekenstein-Hawking entropy of BPS black holes in AdS \cite{Benini:2018ywd}. On the other hand, other solutions can be associated with complex semiclassical saddles of the gravitational path-integral \cite{BeniniToappear, AharonyTalk20-11}. A comprehensive picture of all the contributions from BA solutions and the physics they predict is an interesting open problem.

Even for finite $N$, the complete list of solutions to the BAEs is not known. A subclass of exact solutions was found in \cite{Hosseini:2016cyf, Hong:2018viz}, and we will refer to them as Hong-Liu (HL) solutions. These are discrete solutions, in one-to-one correspondence with subgroups of $\bZ_N \times \bZ_N$ of order $N$, and are parametrized by three integers $\{m,n,r\}$ with $N = mn$ and $r \in \bZ_n$. On the other hand, a very interesting conjecture was put forward in \cite{ArabiArdehali:2019orz}: that there could be a sort of correspondence between some solutions to the BAEs and the vacua of the 4d $\cN=1^*$ theory --- which is $\cN=4$ SYM deformed by $\cN=1$ preserving mass terms --- compactified on $S^1$.
The set of vacua of the $\cN=1^*$ theory on $\bR^{3,1}$ \cite{Donagi:1995cf} (that we will review below) include both discrete massiva vacua, which are easily seen to be in one-to-one correspondence with the HL solutions, as well as massless Coulomb vacua, which give rise to continuous families of vacua upon compactification on $S^1$. Indeed, the authors of \cite{ArabiArdehali:2019orz} gave evidence for the existence of continuous families of solutions to the BAEs. On the other hand, they also pointed out that there are many more solutions to the BAEs (both discrete and continuous) that do not fit into the pattern.

In this Letter we will make the conjectural correspondence sharper, by studying in detail the low-rank cases of gauge groups $SU(2)$ and $SU(3)$.

Our first result (Section \ref{sec: BA formula}) is that, for generic $N$, most of the solutions to the BAEs lead to contributions that cancel out. This means that we can formulate a more restrictive set of equations, that we call ``reduced Bethe Ansatz Equations'', such that only solutions to the latter actually contribute. We then conjecture that there is a one-to-one correspondence between branches of solutions to the reduced BAEs and vacua of the $\cN=1^*$ theory on $\bR^{3,1}$. In particular, isolated solutions to the reduced BAEs are in correspondence with massive vacua of the $\cN=1^*$ theory, while complex $k$-dimensional manifolds of solutions are in correspondence with massless Coulomb vacua with $k$ photons.

We test our conjecture in the cases of $SU(2)$ and $SU(3)$ gauge group. In the case of $SU(2)$ (Section \ref{sec: SU(2)}) we show that the reduced BAEs have only 3 solutions, which indeed are precisely the discrete HL solutions. We then show, by expanding the index order by order in the operator dimension and computing many terms, that those 3 solutions exactly reproduce the superconformal index (as it follows from the integral formula).

In the case of $SU(3)$ (Section \ref{sec: SU(3)}), we are able to analytically solve the reduced BAEs and find all solutions. They comprise the 4 HL solutions, as well as a continuous family which has complex dimension 1 and is made of a single connected component. This is precisely in correspondence with the vacua of the $\cN=1^*$ theory. We also show that, in this case, the HL solutions alone do not reproduce the index, and thus the contribution from the continuous family must be non-trivial. We leave the interesting issue of evaluating such a contribution for future work.

The method we use, in the $SU(3)$ case, to solve the reduced BAEs and to exhibit the structure of discrete and continuous solutions sprouts from the observation that the equations factorize into multiple hyperplanes and curves, at whose intersections lie the solutions. We are hopeful that this method can be generalized to higher $N$.

\textit{Note added}: While this work was completed, its results had been announced in \cite{Rizi:2020}, and we were in the process of writing up this Letter, we received the paper \cite{Lezcano:2021qbj} which has considerable overlap with our work.

\section{Bethe Ansatz formula for the superconformal index}
\label{sec: BA formula}

We study the superconformal index of 4d $\cN=4$ SYM \cite{Romelsberger:2005eg, Kinney:2005ej}, which counts (with sign) operators in short representations (1/16  BPS) of the superconformal algebra, preserving a complex supercharge $\cQ$. Using $\cN=1$ notation, the field content of $\cN=4$ SYM is given by a vector multiplet and three chiral multiplets $X_1, X_2, X_3$, all in the adjoint representation of the gauge group, with superpotential $W = \Tr \bigl( X_1[X_2, X_3] \bigr)$. The R-symmetry is $SU(4)_R$, and we choose the Cartan generators $R_{1,2,3}$ such that each one of them assigns charge 2 to one of the chiral multiplets and charge 0 to the other two. States are further labeled by two angular momenta $J_{1,2}$ with semi-integer eigenvalues, generating rotations of two orthogonal planes in $\bR^4$, and we define the fermion number as $F = 2J_1$. All fields in the theory have integer charges under $R_{1,2,3}$, moreover
\be
F = 2 J_{1,2} = R_{1,2,3} \pmod 2 \;.
\ee
The superconformal index is defined as
\be
\label{SCI}
\cI(p,q,y_1, y_2) = \Tr \biggl[ (-1)^F e^{-\beta\{\cQ, \cQ^\dag\}} \, p^{J_1 + \frac12 R_3} \, q^{J_2 + \frac12 R_3} \, y_1^{\frac12(R_1 - R_3)} \, y_2^{\frac12(R_2 - R_3)} \biggr] \;.
\ee
Here $p,q,y_1, y_2$ are fugacities, and it is convenient to introduce chemical potentials $\sigma, \tau, \Delta_1, \Delta_2$ such that
\be
p = e^{2\pi i \sigma} \;,\qquad\qquad q = e^{2\pi i \tau} \;,\qquad\qquad y_a = e^{2\pi i \Delta_a} \;.
\ee
The index is well-defined for $|p|, |q|<1$. It is a single-valued function of the fugacities, therefore periodic under integer shifts of the chemical potentials, because all states have integer charges with respect to the exponents in (\ref{SCI}). Besides, it will be convenient to introduce an auxiliary fugacity $y_3$ and chemical potential $\Delta_3$ such that%
\footnote{This will restore the permutation symmetry acting on the index $a=1,2,3$ in $R_a$, $y_a$ and $\Delta_a$. Such a symmetry is the Weyl group of the global symmetry $SU(3) \subset SU(4)_R$ that commutes with $\cQ$.}
\be
\Delta_1 + \Delta_2 + \Delta_3 - \sigma - \tau \in \bZ \;.
\ee
By standard arguments \cite{Witten:1982df}, the index only counts states annihilated by $\cQ$ and $\cQ^\dag$ and is therefore independent of $\beta$.

When reading off the (weighted) multiplicities of BPS operators from the index in a ``low-temperature'' expansion, it may be convenient to use the original parametrization of \cite{Kinney:2005ej} in terms of fugacities $(t,y,v_1,v_2)$. They are related to our fugacities by
\be
\label{alternative fugacities}
p = t^3 y \;,\qquad q = t^3/y \;,\qquad y_1 = t^2v_1 \;,\qquad y_2 = t^2v_2 \;,\qquad y_3 = t^2/v_1v_2 \;,
\ee
and recall that $y_3$ is an auxiliary variable defined by $y_1y_2y_3 = pq$. In terms of these fugacities, the index reads
\be
\label{SCI old fugacities}
\cI = \Tr_\text{BPS} (-1)^F \, t^{2(\Delta + J_+)} \, y^{2J_-} \, v_1^{\fq_1} \, v_2^{\fq_2} \;.
\ee
Here $J_\pm = (J_1 \pm J_2)/2$ are the Cartan generators of the $SU(2) \times SU(2)$ Lorentz group, $\Delta$ is the dimension of the operators, the trace is over BPS states --- all and only states satisfying $\Delta = 2J_+ + (R_1 + R_2 + R_3)/2$ --- while $\fq_{1,2} = (R_{1,2}-R_3)/2$ are the Cartan generators of the $SU(3) \subset SU(4)_R$ that commutes with $\cQ$.

\paragraph{Integral formulation.}
The index is independent of the gauge coupling, and thus can be computed exactly \cite{Sundborg:1999ue, Aharony:2003sx, Romelsberger:2005eg, Kinney:2005ej} in terms of a certain contour integral. Specializing to the case of gauge group $SU(N)$, the integral formula reads
\be
\label{integral formula}
\cI_{SU(N)} = \kappa_N \int_{\bT^{N-1}} \cZ(u; \Delta, \sigma, \tau) \, \prod_{j=1}^{N-1} \frac{dz_j}{2\pi i z_j} \;,
\ee
where $z_j = e^{2\pi i u_j}$, and $u_j$ with $j=1,\dots, N$ are gauge holonomies along the Cartan generators. We regard $u_j$ with $j=1, \dots, N-1$ as the independent variables, while the last holonomy, $u_N$, is fixed by the $SU(N)$ constraint
\be
\label{SU(N) constraint}
\sum_{j=1}^N u_j = 0 \;.
\ee
The prefactor is
\be
\kappa_N = \frac1{N!} \, \Bigl( \cI_{U(1)} \Bigr)^{N-1} \;,
\ee
written in terms of the index
\be
\cI_{U(1)}(p,q,y_1,y_2) = (p;p)_\infty (q;q)_\infty \prod_{a=1}^3 \wt\Gamma(\Delta_a; \sigma, \tau)
\ee
of the free $\cN=4$ vector multiplet. Here $(z;q)_\infty$ is the $q$-Pochhammer symbol while $\wt\Gamma(u; \sigma, \tau)$ is the elliptic gamma function, both defined in Appendix~\ref{app: functions}. The integrand is
\be
\cZ(u; \Delta, \sigma, \tau) = \prod_{i\neq j}^N \frac{ \prod_{a=1}^3 \wt\Gamma\bigl( u_{ij} + \Delta_a; \sigma, \tau \bigr) }{ \wt\Gamma\bigl( u_{ij}; \sigma, \tau \bigr) } \;,
\ee
where $u_{ij} = u_i - u_j$. This expression can be slightly simplified using the identity
\be
\prod_{i\neq j}^N \frac1{\wt\Gamma(u_{ij}; \sigma, \tau)} = \prod_{i<j}^N \theta_0(u_{ij}; \sigma) \: \theta_0(u_{ji}; \tau) \;,
\ee
written in terms of the elliptic function $\theta_0$ defined in Appendix~\ref{app: functions}. The contour of integration is along $N-1$ copies of the unit circle, as long as the fugacities satisfy $|pq| < |y_a| < 1$ for all $a=1,2,3$ (otherwise a deformation of the contour is necessary in order to avoid crossing poles).

The superconformal index of the $\cN=4$ SYM theory with gauge group $U(N)$ is related to the one for gauge group $SU(N)$ by the simple relation
\be
\label{U SU relation}
\cI_{U(N)} = \cI_{U(1)} \, \cI_{SU(N)} \;.
\ee

\paragraph{Bethe Ansatz formulation.} In order to compute the large $N$ limit, it is convenient to consider the alternative Bethe Ansatz formula \cite{Closset:2017bse, Benini:2018mlo}. Such a formula applies whenever the angular chemical potentials can be brought, with suitable integer shifts, to have rational ratio,%
\footnote{Such a set is dense in the space of angular chemical potentials.}
$\sigma / \tau \in \bQ$. This means that $\sigma = a\omega$, $\tau = b\omega$ for positive coprime integers $a,b$ and a chemical potential $\omega$ with $\im\omega>0$. Our discussion of the Bethe Ansatz equations will be general, however, for simplicity when computing the index we will restrict to the case of two equal angular fugacities:
\be
p= q \qquad\Leftrightarrow\qquad \sigma = \tau \;.
\ee
In this case the Bethe Ansatz formula simplifies to
\be
\label{BA formula}
\cI(q, y_1, y_2) = \kappa_N \sum_{\hat u \,\in\, \text{BAEs}} \cZ(\hat u; \Delta, \tau) \, H(\hat u; \Delta, \tau)^{-1} \;.
\ee
The sum is over the solution set $\{\hat u\}$ to a system of transcendental equations, dubbed Bethe Ansatz Equations. Defining the $U(N)$ Bethe operators as
\be
\label{BA operators}
Q_j(u; \Delta, \tau) = e^{-2\pi i \sum_{k=1}^N u_{jk}} \prod_{a=1}^3 \prod_{k=1}^N \frac{ \theta_0( u_{kj} + \Delta_a; \tau) }{ \theta_0(u_{jk} + \Delta_a; \tau) }
\ee
for $j=1, \dots, N$, the $SU(N)$ BA equations are given by
\be
\label{BAEs}
1 = \frac{Q_i}{Q_N} \qquad\qquad\text{for}\quad i = 1, \dots, N-1 \;.
\ee
The unknowns are the ``complexified $SU(N)$ holonomies'' $u_i$ living on a torus of modular parameter $\tau$, namely with identifications
\be
u_i \,\sim\, u_i + 1 \,\sim\, u_i + \tau \qquad\qquad\text{for}\quad i=1, \dots, N-1 \;,
\ee
while $u_N$ is fixed by the constraint (\ref{SU(N) constraint}). The BAEs (\ref{BAEs}) are invariant under such shifts.
In fact, a stronger property holds: $Q_i$ are invariant under shifts of the components of the antisymmetric tensor $u_{ij}$ by $1$ or $\tau$, even relaxing the condition that $u_{ij} = u_i - u_j$. This will be used later.
It was proven in \cite{Benini:2018mlo} that only the solutions that are not invariant under any non-trivial element of the Weyl group of $SU(N)$ (namely, only solutions with all $u_i$ different on the torus) actually contribute to the sum in (\ref{BA formula}). In the general case $\sigma/\tau \in \bQ$, the BA formula takes a more complicated form, however the BAEs are the same as in (\ref{BAEs}) with the only difference that $\tau$ is replaced by $\omega$. Therefore the analysis we will make of the BAEs and their solutions will be valid in the general case.

The prefactor $\kappa_N$ and the integrand $\cZ$ in (\ref{BA formula}) are the same as in the integral formula described before. On the other hand, $H$ is a Jacobian defined as
\be
\label{def Jacobian}
H = \det\left[ \frac1{2\pi i} \, \parfrac{\log(Q_i / Q_N)}{u_j} \right]_{i,j=1, \dots, N-1} \;.
\ee
Notice that $Q_i$, $\kappa_N$, $\cZ$ and $H$ are all invariant under integer shifts of $\tau$, $\Delta_1$ and $\Delta_2$, in accord with the fact that the superconformal index (\ref{BA formula}) is a single-valued function of the fugacities.

\paragraph{Reduced Bethe Ansatz equations.} It turns out that not all solutions (faithfully acted upon by the Weyl group) of (\ref{BAEs}) actually contribute to the sum in (\ref{BA formula}). In order to understand this point, notice that $\prod_{i=1}^N Q_i = 1$ identically. It follows that the solutions to the BAEs (\ref{BAEs}) break into $N$ ``sectors'' parametrized by $\lambda \in \bZ_N$ and given by
\be
Q_j = e^{2\pi i \lambda/N} \qquad\qquad\text{for}\quad j = 1, \dots, N \;,
\ee
each sector corresponding to a different integer value of $\lambda = 0, 1, \dots, N-1$ and containing a subset of the solutions. Now, the Bethe operators have the property that they implement on $\cZ$ the shifts of complexified $SU(N)$ holonomies on the torus \cite{Closset:2017bse, Benini:2018mlo}:
\be
\cZ(u- \delta_k \tau) = \frac{Q_k(u)}{Q_N(u)} \, \cZ(u) \qquad\qquad\text{for}\quad k = 1, \dots, N-1 \;,
\ee
where $u - \delta_k \tau = (u_1, \dots, u_k - \tau, \dots, u_{N-1}, u_N+\tau)$ denotes a shift of the $k$-th and $N$-th components of $u$, so as to preserve the $SU(N)$ constraint. On the other hand, if we regard $\cZ(u)$ as function of $N$ independent holonomies $u_i$ of $U(N)$, we find
\be
\label{shift of Z}
\cZ\bigl( u - \bar\delta_k \tau \bigr) = (-1)^{N-1} \, Q_k(u) \, \cZ(u) \qquad\qquad\text{for}\quad k = 1, \dots, N \;,
\ee
where $u - \bar\delta_k \tau = (u_1, \dots, u_k - \tau, \dots, u_N)$ denotes a shift of the $k$-th component only.%
\footnote{The factor $(-1)^{N-1}$ did not appear in \cite{Benini:2018mlo} because that paper only dealt with semi-simple gauge groups.}
The BAEs only depend on the differences $u_{ij}$ and, as already noted, are invariant under shifts of $u_{ij}$ by multiples of $1$ and $\tau$. It follows that one solution for $u_{ij}$ on the torus gives rise to multiple solutions for $u_i$, of the form $u_i = u_i^{(0)} + (\alpha + \beta\tau)/N$ for $i=1, \dots, N-1$ and $u_N = u_N^{(0)} + (1-N)(\alpha + \beta\tau)/N$, where $\alpha,\beta = 0, \dots, N-1$ (some of these solutions could be equivalent up to the Weyl group action). The function $\cZ$, to be evaluated in (\ref{BA formula}) on those solutions, only depends on $u_{ij}$, thus the dependence on $\alpha,\beta$ is the same as if $u_N$ were shifted by $-(\alpha + \beta\tau)$. There is no dependence on $\alpha$, while the dependence on $\beta$ is a phase $\bigl( (-1)^{N-1} Q_N(u) \bigr)^\beta$ described by (\ref{shift of Z}). We conclude that in all but one sector of BA solutions, the sum over $\beta$ leads to a cancelation because the sum of phases vanishes. The exception is the sector in which $(-1)^{N-1} Q_N(u) =1$.

Therefore, in the Bethe Ansatz formula (\ref{BA formula}) we can restrict to BA solutions solely in the sector $\lambda = \frac{N(N-1)}2 \text{ mod } N$, namely to solutions of
\be
\label{reduced BAEs}
Q_i = (-1)^{N-1} \qquad\qquad\text{for}\quad i = 1, \dots, N
\ee
with $Q_i$ given in (\ref{BA operators}), which moreover are faithfully acted upon by the Weyl group. We call these the ``reduced Bethe Ansatz Equations''. Recall that the product of the $N$ equations is identically equal to 1, therefore one of them could be removed from the set. 

\paragraph{Hong-Liu (HL) solutions.} The full set of solutions to (\ref{reduced BAEs}) is not known, however, a large set was found in \cite{Hosseini:2016cyf, Hong:2018viz} and we will refer to them as HL solutions. They are labelled by three positive integers:
\be
\{m,n,r\} \qquad\text{such that}\qquad N = m \cdot n \;,\qquad r \in \bZ_n \;.
\ee
The solutions are
\be
\label{HL solutions u_j}
u_j \equiv u_{\hat\jmath \hat k} = \bar u + \frac{\hat\jmath}m + \frac{\hat k}n \left( \tau + \frac rm \right) \;.
\ee
Here we have decomposed the index $j= 0, \dots, N-1$ into the indices $\hat\jmath = 0, \dots, m-1$ and $\hat k = 0, \dots, n-1$. Moreover, $\bar u$ is a constant chosen in such a way to solve the $SU(N)$ constraint. Since what enters in all formulas are the differences $u_{ij} = u_i - u_j$, to each HL solution is associated a multiplicity that we will discuss below. Notice that the HL solutions are in one-to-one correspondence with subgroups of $\bZ_N \times \bZ_N$ of order $N$, that is with sublattices of index $N$ in generic two-dimensional lattices. It is worth remarking that, surprisingly enough, the HL solutions (\ref{HL solutions u_j}) do not depend on the flavor chemical potentials $\Delta_a$, although the BAEs do.

The BAEs (\ref{reduced BAEs}) are invariant under $SL(2,\bZ)$ modular transformations of the torus, namely under the generators
\be
T : \begin{cases} \tau \mapsto \tau + 1 \\ u \mapsto u \end{cases} \qquad
S : \begin{cases} \tau \mapsto - 1/\tau \\ u \mapsto u/\tau \end{cases} \qquad
C : \begin{cases} \tau \mapsto \tau \\ u \mapsto -u \;. \end{cases}
\ee
It follows that the HL solutions form orbits under $PSL(2,\bZ)$, completely classified by the integer $d = \gcd(m,n,r)$. The action of $PSL(2,\bZ)$ is given by
\be
T: \{m,n,r\} \mapsto \{m, n, r+m\} \;,\qquad
S: \{m,n,r\} \mapsto \left\{ \gcd(n,r) \,,\, \frac{m\, n}{\gcd(n,r)} \,,\, \frac{m(n-r)}{\gcd(n,r)} \right\} \;.
\ee
For given $N$, the total number of HL solutions is given by the divisor function
\be
\sigma_1(N) = \sum_{k | N} k \;.
\ee
For each integer $d$ such that $d^2$ divides $N$, there exist a separate $PSL(2,\bZ)$ orbit of HL solutions generated by $\bigl\{ d, \frac Nd , 0 \bigr\}$, which is isomorphic to the orbit generated by $\{1, N/d^2, 0\}$ in the case of $SU(N/d^2)$. The number of elements in such an orbit is given by $\psi \bigl( N/d^2 \bigr)$, expressed in terms of the Dedekind psi function%
\footnote{Notice that, consistently, $\displaystyle \sum_{d^2|N} \psi\left( \frac{N}{d^2} \right) = \sigma_1(N)$.}
\be
\psi(n) = n \prod_{p|n} \left( 1 + \frac1p \right)
\ee
where the product is over all prime numbers that divide $n$.

The contributions to the BA formula (\ref{BA formula}) from HL solutions connected by the action of $T$ have a simple relation. Indeed, $T$ acts on solutions to the BAEs by $q \to e^{2\pi i} q$, and there is no extra effect of this shift on the summand in  (\ref{BA formula}). This implies that the action of $T$ on the contributions to the BA formula is also given by $q \to e^{2\pi i} q$. In terms of the alternative set of fugacities $(t, v_1, v_2)$ defined in (\ref{alternative fugacities}) (we are setting $y=1$ here), this is
\be
\label{T action on contributions}
t \,\to\, e^{\frac{2\pi i}3} \, t \;,\qquad\qquad v_a \,\to\, e^{\frac{2\pi i}3} \, v_a \;.
\ee
Since the action of $T$ on HL solutions forms finite cyclic orbits, this can pose constraints on the contributions from those solutions (although notice that, as opposed to the full index, the separate contributions to (\ref{BA formula}) in general are not single-valued functions of the fugacities). We will see this in the examples in Section~\ref{sec: SU(2)} and \ref{sec: SU(3)}.

\paragraph{Multiplicities.}
From each solution in terms of $u_{ij}$ on the torus, one gets up to $N^2$ solutions in terms of $u_i$, related by a shift of the ``center of mass'' of the first $N-1$ components. Some of those, however, could be equivalent up to the Weyl group action, and thus the exact multiplicity should be determined by a case-by-case analysis. In the case of the HL solutions, one easily verifies that each solution has multiplicity $N$, namely it gives rise to $N$ inequivalent solutions in terms of $u_i$ that cannot be identified by the action of the Weyl group.

Besides, since we are only interested in BA solutions that are not fixed by any non-trivial element of the Weyl group, there is an obvious multiplicity by $N!$ related to the action of the Weyl group. Therefore, each HL solution has total multiplicity $N \cdot N!$.

\subsection{Correspondence with the \matht{\cN=1^*} theory}

It has been pointed out in \cite{ArabiArdehali:2019orz} that the Hong-Liu ones are not the only solutions to the BAEs. Furthermore, by a combination of analytical and numerical work, evidence was given that for $N \geq 3$ there exist continuous families of solutions. Very interestingly, the authors of \cite{ArabiArdehali:2019orz} put forward the idea of a correspondence between a subset of the solutions to the BAEs and the vacua of the $\cN=1^*$ $SU(N)$ theory on $S^1$. This, in particular, allowed them to predict the appearance of continuous branches of BA solutions for various values of $N$. On the other hand, it was already noticed in that work that, even for $N=2$, the BAEs admit solutions that do not have a counterpart in the $\cN=1^*$ theory,%
\footnote{The compactification on $S^1$ poses another problem, if one wants to make the correspondence precise. Indeed, certain gapped vacua of the $\cN=1^*$ theory on $\bR^{3,1}$ give rise to multiple vacua on $\bR^{2,1} \times S^1$, due to the presence of a residual discrete gauge symmetry \cite{Aharony:2013hda, Bourget:2016yhy}. For instance, the $SU(2)$ theory has 3 vacua on $\bR^{3,1}$, but 4 vacua on $\bR^{2,1} \times S^1$, while the reduced BAEs have 3 isolated solutions. We are grateful to Jan Troost for making us appreciate this point.}
therefore the problem of understanding in which terms the correspondence could be correct and precise remained open.

In this Letter we present a sharper version of the conjecture: we propose that there is a one-to-one correspondence between branches of solutions to the reduced Bethe Ansatz equations, and vacua of the $\cN=1^*$ $SU(N)$ theory on $\bR^{3,1}$. In particular, isolated solutions to the reduced BAEs are in correspondence with massive vacua of the $\cN=1^*$ theory, while complex $k$-dimensional manifolds of solutions are in correspondence with massless Coulomb vacua with $k$ photons.
If the conjecture is correct, it implies that the HL solutions are the only discrete solutions to the reduced BAEs. It also implies a precise characterization of the number of continuous families and their dimensionality, for each value of $N$. In Sections~\ref{sec: SU(2)} and \ref{sec: SU(3)} we provide some evidence of the conjecture.

Let us review the $\cN=1^*$ $SU(N)$ theory, which is obtained from $\cN=4$ SYM by an $\cN=1$ preserving mass deformation
\be
W_\text{def} = \sum_{a=1}^3 \frac{m_a}2 \, X_a^2 \;.
\ee
Its vacua have been analyzed in \cite{Donagi:1995cf} (see also \cite{Bourget:2015lua, Bourget:2016yhy}).
F- and D-term equations reduce to the classification of isomorphism classes of homomorphisms $\rho: \fsu(2) \to \fsu(N)$ modulo $SL(N,\bC)$, which correspond to partitions of $N$. Since the chiral superfields $X_a$ all have bare masses, possible massless fields come from the gauge sector. At the classical level, the $SU(N)$ gauge group is broken to the subgroup $H$ that commutes with the image of $\rho$. Quantum mechanically, all simple $SU(n)$ factors of $H$ confine leading to a number $n$ of massive sectors; on the other hand, if $H$ contains $U(1)$ factors, then one obtains a massless Coulomb phase.

The net result can be summarized as follows. Consider all integer partitions of $N$. For each partition, let $n_j$ be the number of times the integer $j$ appears in it, and so label the partition by a string
\be
(n_1, \dots, n_j) \qquad\text{such that}\qquad \sum_{j=1}^N j\, n_j = N \;.
\ee
Then the subgroup $H$ is given by%
\footnote{More precisely, $H$ can contain extra discrete factors \cite{Bourget:2016yhy}. The latter would affect the number of vacua upon compactification on $S^1$.}
\be
H = \Biggl[ \; \prod_{j \text{ s.t. } n_j \neq 0}^N U(n_j) \, \Biggr] / U(1) \;.
\ee
Thus, such a partition contributes to the moduli space with a certain number of vacua that can be either gapped or contain massless photons:
\begin{itemize}
\item $\text{number of photons} = ( \text{number of non-zero $n_j$'s} ) - 1$ (these are isolated gapped vacua if the number is zero, or Coulomb branch vacua if it is bigger than zero);
\item number of vacua = product of non-zero $n_j$'s.
\end{itemize}
It turns out \cite{Donagi:1995cf} that the massive vacua are in one-to-one correspondence with subgroups of order $N$ of $\bZ_N \times \bZ_N$, and thus in correspondence with the HL solutions to the BAEs. On the other hand, we put each Coulomb vacuum in correspondence with a continuous family of solutions to the BAEs, with complex dimension equal to the number of photons.

\section{The case of \matht{SU(2)}}
\label{sec: SU(2)}

Let us analyze in detail the case of gauge group $SU(2)$. Defining $u \equiv u_{21} = u_2 - u_1$, the refined Bethe Ansatz equation (\ref{reduced BAEs}) can be rewritten as
\be
\label{reduced BAE SU(2)}
\frac{\theta_{1}(\Delta_{1} + u ) \, \theta_{1} ( \Delta_{2} + u ) \, \theta_{1} ( -\Delta_{1}-\Delta_{2} + u ) }{ \theta_{1} ( \Delta_{1} - u ) \, \theta_{1} ( \Delta_{2} - u ) \, \theta_{1} ( -\Delta_{1}-\Delta_{2} - u ) } = -1 \;,
\ee
in terms of the standard Jacobi theta function $\theta_1$ (see Appendix~\ref{app: functions}). For the sake of clarity, here and in the following we leave the dependence of the Jacobi theta functions $\theta_r$ on $\tau$ implicit. Let us define the function
\be
\label{def f}
f(u; \Delta, \tau) = \theta_1(u) \, \theta_1( \Delta_1 + u) \, \theta_1 ( \Delta_2 + u) \, \theta_1 ( -\Delta_1 - \Delta_2 + u) \;.
\ee
Then, multiplying and dividing (\ref{reduced BAE SU(2)}) by $\theta_1(u)$ and using that $\theta_1$ is an odd function of $u$ (while $\theta_{2,3,4}$ are even functions of $u$), we can rewrite the reduced BAE as
\be
\frac{f(u; \Delta, \tau) - f (-u; \Delta, \tau)}{f(-u; \Delta, \tau)} = 0 \;.
\ee
Using standard identities among the theta functions, see for instance \cite{Kharchev:2015}, we obtain
\be
\label{theta relation SU(2)}
f(u; \Delta, \tau) = c_1 \, \theta_1(2u) +  c_2 \, \theta_2(2u) + c_3 \, \theta_3(2u) + c_4 \, \theta_4(2u)
\ee
where the coefficients are given by
\bea
\label{coefficients SU(2)}
c_1 &= -\frac{1}{2} \, \theta_1(\Delta_1) \, \theta_1(\Delta_2) \, \theta_{1}(\Delta_{1}+\Delta_{2})\\
c_2 &= -\frac{1}{2} \, \theta_2(\Delta_1) \, \theta_2(\Delta_2) \, \theta_{2}(\Delta_{1}+\Delta_{2}) \\
c_3 &= + \frac{1}{2} \, \theta_3(\Delta_1) \, \theta_3 (\Delta_2) \, \theta_3(\Delta_1+\Delta_2) \\
c_4 &= -\frac{1}{2} \, \theta_4(\Delta_1) \, \theta_4(\Delta_2) \, \theta_{4}(\Delta_{1}+\Delta_{2}) \;.
\eea
From the parity properties of the theta functions, the reduced BAE  becomes
\be
\frac{c_1(\Delta, \tau) \, \theta_1(2 u) }{ f(-u; \Delta,\tau)} = 0 \;.
\ee
For generic values of $\Delta_a$ and $\tau$, the coefficient $c_1$ is non-zero and thus the full set of solutions is given by $u = (m+n\tau)/2$ for integers $m,n$ that are not both even.%
\footnote{When $m,n$ are both even, the zero of $\theta_1$ in the numerator cancels with the zero of $f$ in the denominator.}
On the torus there are 3 solutions:%
\footnote{Notice also that $u=0$ would lead to configurations $(u_1,u_2)$ that are fixed by the non-trivial element of the Weyl group, and thus would have to be excluded even if it was a solution.}
\be
u = \frac{1}{2} \,,\; \frac{\tau}{2} \,,\; \frac{\tau+1}{2} \;.
\ee
These are precisely the 3 Hong-Liu solutions $\{2,1,0\}$, $\{1,2,0\}$, $\{1,2,1\}$, respectively, as it follows from (\ref{HL solutions u_j}). Therefore, for $N=2$, the HL solutions exhaust the full set of solutions to the reduced BAEs, in agreement with the proposed correspondence with vacua of the $\cN=1^*$ theory.

\paragraph{Evaluation of the index.}
According to the BA formula (in the special case $\sigma=\tau$ for simplicity), we have the identity
\be
\label{identity for SU(2)}
\cI_{SU(2)}(q,y_1,y_2) = \cI_{SU(2)}^{\{2,1,0\}} + \cI_{SU(2)}^{\{1,2,0\}} + \cI_{SU(2)}^{\{1,2,1\}} \;.
\ee
On the left-hand side of this identity, $\cI_{SU(2)}$ is full index computed with the integral formula (\ref{integral formula}). On the right-hand side, instead, each of the three terms $\cI_{SU(2)}^{\{m,n,r\}}$ is the contribution to the Bethe Ansatz formula (\ref{BA formula}) from one of the HL solutions:
\be
\label{BA contributions}
\cI_{SU(N)}^{\{m,n,r\}} = N \!\cdot\! N! \, \kappa_N \, \cZ(\hat u; \Delta, \tau) \, H(\hat u; \Delta, \tau)^{-1} \, \Big|_{\hat u \in \{m,n,r\}} \;,
\ee
where $\hat u_j$ are as in the HL solution $\{m,n,r\}$ given in (\ref{HL solutions u_j}), while the factor $N \cdot N!$ comes from the multiplicity. We have verified (\ref{identity for SU(2)}) numerically for many values of the fugacities. This confirms that, for $N=2$, the contributions from the three HL solutions exactly reproduce the superconformal index.

To gain a more interesting physical understanding, we can perform a ``low-temperature'' expansion of both sides of (\ref{identity for SU(2)}), corresponding to an expansion of the index into BPS operators, order by order in the operator dimension and starting from the identity. In order to do that, it is convenient to use the set of fugacities $(t,y, v_1, v_2)$ in (\ref{SCI old fugacities}). The restriction to $p=q$ corresponds to $y=1$. On the other hand, the dependence on $v_1,v_2$ organizes into characters of $SU(3)$ that we indicate by $\chi_\text{dim}$. For instance
\bea
\chi_\mathbf{1} &= 1 \\
\chi_\mathbf{3} &= v_1 + v_2 + v_1^{-1} v_2^{-1} \\
\chi_\mathbf{6} &= v_1^2 + v_2^2 + v_1v_2 + v_1^{-1} + v_2^{-1} + v_1^{-2} v_2^{-2} \;,
\eea
and so on. With computer assistance, we perform a Taylor or Laurent expansion of both sides of (\ref{identity for SU(2)}) around $t=0$.

Expanding the integral expression, left-hand side of (\ref{identity for SU(2)}), we obtain the following. Switching off also the flavor fugacities ($v_1 = v_2 = 1$) we find%
\footnote{A much longer list of coefficients can be found in \cite{Murthy:2020rbd} for gauge group $U(2)$, and then the coefficients for $SU(2)$ can be derived using (\ref{U SU relation}). Switching off flavor fugacities, one finds
\be
\cI_{U(1)} = 1 + 3t^2 - 2t^3 + 3t^4 +6 t^7 - 6 t^8 + 12 t^{10} - 18 t^{11} + 27 t^{12} - 12 t^{13} - 27 t^{14} + \cO(t^{15}) \;.
\ee}
\be
\label{SU(2) integral exp t}
\cI_{SU(2)} = 1+ 6 \, t^{4} -6 \, t^{5} - 7 \, t^{6}+18 \, t^{7} + 6 \, t^8 - 36 \, t^{9} + 6 \, t^{10} + 84 \, t^{11} - 80 \, t^{12} -132 \, t^{13} + 309 \, t^{14}  + \cO(t^{15}) .
\ee
Notice that the expansion only contains integer powers of $t$, and the coefficients are well-defined because the number of operators with bounded dimension is finite. Including the flavor fugacities we find
\bea
\label{SU(2) integral exp v1v2}
\cI_{SU(2)} &= 1+\chi_\rep{6} t^4 - 2 \chi_\rep{3} t^5 + \bigl( 1-\chi_\rep{8} \bigr) t^6 + 2 \bigl( \chi_\rep{\bar{3}} + \chi_\rep{6} \bigr) t^7 +\bigl( \chi_\rep{15} - 3 \chi_\rep{3} \bigr) t^8 \\
&\qquad {} - 2 \bigl( \chi_\rep{10} + \chi_\rep{8} \bigr) t^9 + \bigl( 3 \chi_\rep{6} + 4 \chi_\rep{\bar{3}} - \chi_\rep{24} \bigr) t^{10} + \cO(t^{11} ) \;.
\eea
In terms of Dynkin labels: $\rep{10} = [3,0]$, $\rep{15} = [4,0]$, $\rep{15'} = [2,1]$, $\rep{21} = [5,0]$ and $\rep{24} = [3,1]$.

Similarly, we expand the BA contributions (\ref{BA contributions}) for each one of the three HL solutions. Suppressing flavor fugacities, we find
\bea
\label{SU(2) BA exp t}
\cI_{SU(2)}^{\{2,1,0\}} &= - 2 - 12 t - 42 t^2 - 124 t^3 - 348 t^4 - 900 t^5 - 2142 t^6 - 4860 t^7 - 10644 t^8 + \cO(t^9) \\ 
\cI_{SU(2)}^{\{1,2,0\}} &= - \frac{1}{4} t^{-\frac32} - \frac{3}{4} t^{-\frac12} + \frac{3}{2} - 3 t^{\frac12} + 6 t -\frac{43}{4} t^{\frac32} + 21 t^2 - \frac{153}{4} t^{\frac52} + 62 t^3 - 105 t^{\frac72} \\
&\quad\; + 177 t^4 - \frac{1131}{4} t^{\frac92} + 447 t^5 - \frac{2775}{4} t^{\frac{11}2} + \frac{2135}{2} t^6 - 1635 t^{\frac{13}2} + 2439t^7+\cO\bigl( t^{\frac{15}2} \bigr) \\
\cI_{SU(2)}^{\{1,2,1\}} &= + \frac{1}{4} t^{-\frac32} + \frac{3}{4} t^{-\frac12} + \frac{3}{2} + 3 t^{\frac12} + 6 t +\frac{43}{4}t^{\frac32} + 21 t^2 + \frac{153}{4} t^{\frac52} + 62 t^3 + 105 t^{\frac72} \\
&\quad\; + 177 t^4 + \frac{1131}{4} t^{\frac92} + 447 t^5 +\frac{2775}{4} t^{\frac{11}2} + \frac{2135}{2} t^6 + 1635 t^{\frac{13}2} + 2439t^7 + \cO\bigl( t^{\frac{15}2} \bigr) \;.
\eea
Interestingly, the contributions from the two $\{1,2,r\}$ solutions contain many unwanted features: fractional coefficients, and fractional as well as negative powers of $t$. However the unwanted powers of $t$ cancel out in the sum, and the coefficients sum up to integers. We have verified that the sum of the three contributions above exactly matches the expansion (\ref{SU(2) integral exp t}) of the integral expression, up to order $\cO(t^{40})$. Including the flavor fugacities we find
\newcommand{\tenbar}{\mathbf{\bar1\!\!\bar{\phantom{1}}\!\bar0}}
\newcommand{\fifbar}{\mathbf{\bar1\!\!\bar{\phantom{1}}\!\bar5}}
\newcommand{\fifbarprime}{\mathbf{\bar1\!\!\bar{\phantom{1}}\!\bar5'}}
\newcommand{\twonebar}{\mathbf{\bar2\!\!\bar{\phantom{1}}\!\bar1}}
\newcommand{\twfourbar}{\mathbf{\bar2\!\!\bar{\phantom{1}}\!\bar4}}
\newcommand{\tweightbar}{\mathbf{\bar2\!\!\bar{\phantom{1}}\!\bar8}}
\newcommand{\thfivebar}{\mathbf{\bar3\!\!\bar{\phantom{1}}\!\bar5}}
\begin{align}
 \cI_{SU(2)}^{\{2,1,0\}} &= - 2 - 4 \chi_\rep{\bar3} t - 2 \bigl( 3 \chi_\rep{3} + 2 \chi_\rep{\bar6} \bigr) t^2 - 4 \bigl( \chi_\tenbar + 2 \chi_\rep{8} + 5 \bigr) t^3 \\
 &\qquad - 4 \bigl( \chi_\fifbar + 2 \chi_\fifbarprime + 2 \chi_\rep{6} + 10 \chi_\rep{\bar3} \bigr) t^4 +\cO(t^5) \nn \\
\cI_{SU(2)}^{\{1,2,0\}} &= -\frac{1}{4} t^{-\frac32} - \frac{1}{4} \chi_\rep{\bar3} t^{-\frac12} + \frac{3}{2} - \frac{1}{4} \bigl( \chi_\rep{\bar6} + 2 \chi_\rep{3} \bigr) t^{\frac12} + 2 \chi_\rep{\bar3} t - \frac{1}{4} \bigl( \chi_\tenbar + 2 \chi_\rep{8} + 17 \bigr) t^{\frac32} \nn \\
&\;\; + \bigl( 2 \chi_\rep{\bar6} + 3 \chi_\rep{3} \bigr) t^2 - \frac{1}{4} \bigl( \chi_\fifbar + 2 \chi_\rep{15'} + 3 \chi_\rep{6} + 30 \chi_\rep{\bar3} \bigr) t^{\frac52} + 2 \bigl( \chi_\tenbar + 2 \chi_\rep{8} + 5 \bigr) t^3 \nn \\
&\;\; - \frac{1}{4} \bigl( \chi_\twonebar + 2 \chi_\twfourbar + 3 \chi_\rep{15'} + 31 \chi_\rep{\bar6} + 40 \chi_\rep{3} \bigr) t^{\frac72} + \bigl( 2 \chi_\fifbar + 4 \chi_\fifbarprime + \tfrac92 \chi_\rep{6} + 20 \chi_\rep{\bar3} \bigr) t^4 + \cO \bigl( t^{\frac92} \bigr) \nn \\
\cI_{SU(2)}^{\{1,2,1\}} &= + \frac{1}{4} t^{-\frac32} + \frac{1}{4} \chi_\rep{\bar3} t^{-\frac12} + \frac{3}{2} + \frac{1}{4} \bigl( \chi_\rep{\bar6} + 2 \chi_\rep{3} \bigr) t^{\frac12} + 2 \chi_\rep{\bar3} t + \frac{1}{4} \bigl( \chi_\tenbar + 2 \chi_\rep{8} + 17 \bigr) t^{\frac32} \nn \\
&\;\; + \bigl( 2 \chi_\rep{\bar6} + 3 \chi_\rep{3} \bigr) t^2 + \frac{1}{4} \bigl( \chi_\fifbar + 2 \chi_\rep{15'} + 3 \chi_\rep{6} + 30 \chi_\rep{\bar3} \bigr) t^{\frac52} + 2 \bigl( \chi_\tenbar + 2 \chi_\rep{8} + 5 \bigr) t^3 \nn \\
&\;\; + \frac{1}{4} \bigl( \chi_\twonebar + 2 \chi_\twfourbar + 3 \chi_\rep{15'} + 31 \chi_\rep{\bar6} + 40 \chi_\rep{3} \bigr) t^{\frac72} + \bigl( 2 \chi_\fifbar + 4 \chi_\fifbarprime + \tfrac92 \chi_\rep{6} + 20 \chi_\rep{\bar3} \bigr) t^4 + \cO \bigl( t^{\frac92} \bigr) \nn
\end{align}
Notice that the action of $T$ on these expressions is correctly given by (\ref{T action on contributions}). In particular, $\cI_{SU(2)}^{\{2,1,0\}}$ is invariant, while $\cI_{SU(2)}^{\{1,2,1\}}$ is obtained from $\cI_{SU(2)}^{\{1,2,0\}}$ (and viceversa). In both cases, the fact that $T$ has finite cyclic orbits and that only integer or half-integer powers of $t$ appear, implies that the triality of the $SU(3)$ characters is correlated with the power of $t$. We have verified that the sum of the three contributions above reproduces the expansion (\ref{SU(2) integral exp v1v2}) of the integral expression, up to order $\cO(t^{10})$.

\section{The case of \matht{SU(3)}}
\label{sec: SU(3)}

We consider now the case of gauge group $SU(3)$. We define $u \equiv u_{21}$ and $v \equiv u_{31}$. The Weyl group $S_3$ of the $SU(3)$ gauge group permutes the triplet $(u_1, u_2, u_3)$ while it acts in the following way on $(u,v)$:
\bea
s_{12}: (u,v) &\mapsto (-u, v-u) \qquad\qquad& s_R: (u,v) &\mapsto (v-u, -u) \\
s_{13}: (u,v) &\mapsto (u-v, -v) \qquad\qquad& s_L: (u,v) &\mapsto (-v, u-v) \\
s_{23}: (u,v) &\mapsto (v,u) \;.
\eea
The three reduced BAEs (\ref{reduced BAEs}) are mapped one into the others by the action of the Weyl group, therefore let us discuss how to manipulate the first of them. Introducing the function $f$ defined in (\ref{def f}), we can recast that equation in the form%
\footnote{Notice that the first reduced BAE concerns the antisymmetric part of the product of two $f$'s. This generalizes to any $N$: the solutions to the first reduced BAE can always be written as the zeros of the antisymmetric part of the product of $N-1$ functions $f$ (while the other ones are obtained by the action of the Weyl group).}
\be
0 = Q_1(u,v; \Delta,\tau) -1 = \frac{ f(u; \Delta, \tau) \, f(v; \Delta,\tau) - f(-u; \Delta, \tau) \, f(-v; \Delta, \tau) }{ f(-u; \Delta, \tau) \, f(-v; \Delta, \tau) } \;.
\ee
Using the identity (\ref{theta relation SU(2)}), the numerator becomes
\be
\text{num} = \biggl[ c_1 \, \theta_1(2u) \sum_{k=2,3,4} \! c_k \, \theta_k(2v) \biggr] + \biggl[ c_1 \, \theta_1(2v) \sum_{k=2,3,4} \! c_k \, \theta_k(2u) \biggr] \;.
\ee
Collecting terms with the same coefficient $c_i$ and using the identities (\ref{more theta identities}), we rewrite it as
\begin{align}
\text{num} &= \frac{c_1 \, \theta_1(u+v) }{ \theta_2(0) \, \theta_3(0) \theta_4(0) } \Biggl[ [3] \, \theta_3(u+v) \, \theta_2(u-v) \, \theta_4(u-v) \\
&\hspace{2cm} {} - [4] \, \theta_4(u+v) \, \theta_2(u-v) \, \theta_3(u-v) - [2] \, \theta_2(u+v) \, \theta_3(u-v) \, \theta_4(u-v) \Biggr] \;, \nn
\end{align}
where we defined $[r] = \theta_r(0)\, \theta_r(\Delta_1) \, \theta_r(\Delta_2) \, \theta_r(\Delta_1 + \Delta_2)$. They obey $[3] = [4] + [2]$ \cite{Kharchev:2015}, then using (\ref{more theta identities}) again we obtain
\begin{align}
\text{num} &= \frac{2 \, c_1 \, \theta_1(u+v) \, \theta_1(u) \, \theta_1(v) }{ \theta_2(0)^2 \, \theta_3(0)^2 \, \theta_4(0)^2 } \times {} \\
&\hspace{2cm} {} \times \biggl[ [2] \, \theta_4(0) \, \theta_4(u) \, \theta_4(v) \, \theta_4(u-v) - [4] \, \theta_2(0) \, \theta_2(u) \, \theta_2(v) \, \theta_2(u-v) \biggr] \;. \nn
\end{align}
This allows us to rewrite the first reduced BAE as
\be
\label{expr 1}
0 = Q_1 - 1 = \frac{2\, c_1 \, \theta_1(u+v) \, h(u,v; \Delta, \tau) }{ \theta_2(0) \, \theta_3(0)^2 \, \theta_4(0) \prod_{\Delta \in \{\Delta_1, \Delta_2, -\Delta_1 - \Delta_2\}} \theta_1(\Delta -u) \, \theta_1(\Delta - v) }
\ee
where we defined the function
\be
h(u,v;\Delta,\tau) = \theta_2(\Delta_1) \, \theta_2(\Delta_2) \, \theta_2(\Delta_1 + \Delta_2) \, \theta_4(u) \, \theta_4(v) \, \theta_4(u-v) - \bigl( 2 \leftrightarrow 4 \bigr) \;.
\ee
Crucially, this function is invariant under the gauge Weyl group.
The second reduced BAE can be obtained by acting with the gauge Weyl group on the first one. Since the denominator of (\ref{expr 1}) does not have poles, the reduced BAEs can be brought to the factorized form
\be
\label{SU(3) BAEs simplified}
\left\{ \begin{aligned}
0 &= \theta_1(u+v) \, h(u,v; \Delta, \tau) \\
0 &= \theta_1(2u-v) \, h(u,v; \Delta, \tau) \;.
\end{aligned} \right.
\ee
From here we clearly see the structure of the solutions.

First, there are discrete solutions that follow from solving
\be
0 = \theta_1(u+v) = \theta_1(2u-v) \;.
\ee
These equations represent hyperplanes inside $T^2$, and the discrete solutions are at the intersections of two hyperplanes. There are four solutions on $T^2$, up to identifying those that are related by the Weyl group action and dropping those that are fixed by some non-trivial element of the Weyl group (\ie, solutions in which either $u$ or $v$ vanish):
\be
(u,v) = \left( \frac13, \frac23 \right) ,\; \left( \frac\tau3, \frac{2\tau}3 \right) ,\; \left( \frac{\tau + 1}3, \frac{2\tau + 2}3 \right) ,\; \left( \frac{\tau+2}3, \frac{2\tau+1}3 \right) .
\ee
These are precisely the 4 HL solutions $\{3,1,0\}$, $\{1,3,0\}$, $\{1,3,1\}$, $\{1,3,2\}$, respectively, according to (\ref{HL solutions u_j}).

Second, there is a continuous family of solutions obtained by solving the single equation
\be
\label{SU(3) curve}
0 = h(u,v; \Delta,\tau) \;.
\ee
We will analyze some properties of these solutions below, but we already see that the continuous family has complex dimension 1. The solutions to (\ref{SU(3) curve}), as opposed to the HL solutions, depend on the flavor fugacities $\Delta_a$. There are however six special points on the curve,%
\footnote{These points have been studied in \cite{Dorey:1999sj} in the context of the $\cN=1^*$ theory on $S^1$. In the context of the Bethe Ansatz equations, they have been noticed in \cite{ArabiArdehali:2019orz}.}
all related by the action of the $S_3$ Weyl group, that do not depend on $\Delta_a$:
\be
(u,v) = \left( \frac12 , \frac\tau2 \right) \;.
\ee
On these points, the action of $PSL(2,\bZ)$ reduces to the action of $S_3$. These are smooth points, which do not seem to have any other special property.

\paragraph{Evaluation of the index.}
The Bethe Ansatz formula (\ref{BA formula}) was derived in \cite{Closset:2017bse, Benini:2018mlo} assuming that all solutions to the BAEs are isolated. It is then clear that, in the presence of continuous families of solutions as it happens in this case, the formula has to be modified somehow in order to correctly evaluate the contribution of those families. We could hope that, for some reason, continuous families do not contribute. Let us test this possibility, by comparing the result of the integral formula (\ref{integral formula}) with that of the BA formula (\ref{BA formula}), in which the sum is restricted to isolated solutions to the BAEs (here the HL solutions), in the low-temperature limit.

The expansion of the integral formula (\ref{integral formula}), setting the flavor fugacities $v_1 = v_2=1$, gives
\be
\cI_{SU(3)} = 1+ 6 \, t^4 - 6 \, t^5 + 3 \, t^6 + 6 \, t^7 - 16 \, t^9 + 27 \, t^{10} + 18 \, t^{11} - 87 \, t^{12} + 96 \, t^{13} + 54 \, t^{14} + \cO(t^{15}) ,
\ee
while switching on the $SU(3)$ fugacities it gives
\begin{align}
\cI_{SU(3)} &= 1+ \chi_\rep{6} t^4 - 2\chi_\rep{3} t^5 + \bigl( \chi_\rep{10} - \chi_\rep{8} + 1 \bigr) t^6 + 2 \chi_\rep{\bar3} t^7 + \bigl( \chi_\rep{15} - \chi_\rep{15'} + \chi_\rep{\bar6} - 2 \chi_\rep{3} \bigr) t^8 \\
&\quad - 2 \chi_\rep{8} t^9 + \bigl( \chi_\rep{21} - \chi_\fifbarprime + \chi_\rep{6} + 5 \chi_\rep{\bar3} \bigr) t^{10} + \bigl( 2 \chi_\rep{15'} - 2 \chi_\rep{15} + 4 \chi_\rep{\bar6} - 2\chi_\rep{3} \bigr) t^{11} + \cO(t^{12}) \;. \nn
\end{align}
The expansion of the contributions (\ref{BA contributions}) of the HL solutions to the BA formula gives:
\begin{align}
\cI_{SU(3)}^{\{3,1,0\}} &= 3 + 6 \chi_\rep{\bar3} t + 3 \bigl( 5\chi_\rep{\bar6} - 2\chi_\rep{3} \bigr) t^2 + 6 \bigl( 3 \chi_\tenbar + 2 \chi_\rep{8} - 6 \bigr) t^3 \nn \\
&\;\; + 3 \bigl( 10 \chi_\fifbar - \chi_\fifbarprime + 24 \chi_\rep{6} -13 \chi_\rep{\bar3} \bigr) t^4 + 6 \bigl( 5 \chi_\twonebar + 6 \chi_\twfourbar + 8 \chi_\rep{15'} - 22 \chi_\rep{\bar6} + 30 \chi_\rep{3} \bigr) t^5 \nn \\
&\;\; + 3 \bigl( 15 \chi_\tweightbar - \chi_\thfivebar + 16 \chi_\twfourbar + 71( \chi_\rep{27} + \chi_\rep{24} ) - 99 \chi_\tenbar - 69 \chi_\rep{10} - 85 \chi_\rep{8} + 164 \bigr) t^6 + \cO(t^7) \nn \\
\cI_{SU(3)}^{\{1,3,0\}} &= \frac{ \chi_\rep3 - \chi_\rep{\bar3} }{ 9 \bigl( \chi_\rep{6} - 6 \chi_\rep{3} + \chi_\rep{\bar3} + 9 \bigr)} \, t^{-4} \nn \\
&\;\; + \frac{ 2 \bigl( \chi_\rep{15'} - \chi_\fifbarprime + 6 \chi_\rep{8} - 7 \chi_\rep{6} + \chi_\rep{\bar6} + 11 \chi_\rep{3} - 17 \chi_\rep{\bar3} + 6 \bigr) }{ 9 \bigl( \chi_\rep{10} + 2 \chi_\rep{8} - 9 \chi_\rep{6} - 9 \chi_\rep{\bar3} + 27 \chi_\rep{3} - 26 \bigr) } \, t^{-3} + \cO(t^{-2}) \\
\cI_{SU(3)}^{\{1,3,1\}} &= \frac{ \chi_\rep{\bar3} + e^{\pi i/3} \chi_\rep{3}  }{ 9 \bigl( 6 \chi_\rep{3} + e^{-\pi i /3} ( \chi_\rep{6} + \chi_\rep{\bar3}) + 9 e^{\pi i /3} \bigr) } \, t^{-4} \nn \\
&\;\; - \frac{ 2 \bigl( \chi_\rep{15'} + \chi_\rep{\bar6} + 11 \chi_\rep{3} - e^{\pi i /3} (6\chi_\rep{8} + 6) + e^{-\pi i /3} (\chi_\fifbarprime + 7 \chi_\rep{6} + 17 \chi_\rep{\bar3} ) \bigr) }{ 9 \bigl( e^{\pi i /3} ( \chi_\rep{10} + 2 \chi_\rep{8} - 26) - 9 e^{- \pi i/3} ( \chi_\rep{6} + \chi_\rep{\bar3}) - 27\chi_\rep{3} \bigr) } \, t^{-3} + \cO(t^{-2}) \nn 
\end{align}
while $\cI_{SU(3)}^{\{1,3,2\}}$ is obtained from $\cI_{SU(3)}^{\{1,3,1\}}$ by taking the complex conjugate of all coefficients. The action of $T$ on these expressions is given by (\ref{T action on contributions}). Notice that $\cI_{SU(3)}^{\{1,3,r\}}$ does not have a well-defined limit as $v_{1,2} \to e^{-\frac{2\pi i}3 r}$, for $r=0,1,2$.

One immediately sees that the sum of the contributions from the 4 HL solutions does not reproduce the index. In particular, there remain negative powers of $t$, fractional coefficients and characters in the denominators. This implies that the contribution from the continuous family of solutions is necessary in order to correctly reproduce the index. It is an open issue how to compute the contribution from the continuous family, to which we hope to return in future work.

\subsection{The continuous family of solutions}

Let us study in more detail the curve $h(u,v; \Delta, \tau) = 0$, for generic values of the flavor chemical potentials $\Delta_a$.%
\footnote{In the case of the $SU(3)$ $\cN=1^*$ theory on $S^1$, the continuous family of vacua has been studied in \cite{Bourget:2015upj}.}
We rewrite the equation as
\be
\label{eq:equationcont}
\frac{ \theta_4(u) \, \theta_4(v) \, \theta_4(u-v) }{ \theta_2(u) \, \theta_2(v) \, \theta_2(u-v) } = \frac{ \theta_4(\Delta_1) \, \theta_4(\Delta_2) \, \theta_4(\Delta_1 + \Delta_2) }{ \theta_2(\Delta_1) \, \theta_2(\Delta_2) \, \theta_2(\Delta_1 + \Delta_2) } \;.
\ee
If we regard the left-hand side as a function of $u$ (for fixed $v$), we see that it is doubly-periodic with periods $1$ and $\tau$, with two simple poles at $u=\frac12$ and $u=v+\frac12$, and two simple zeros at $u=\frac\tau2$ and $u = v + \frac\tau2$, \ie, it is an elliptic function of order 2. This means that it gives two values $u_\pm(v)$ of $u$ for each value of $v$, and they satisfy $u_+ + u_- = v \pmod{\bZ + \bZ\tau}$. In other words we have a branched double cover of the torus.

The equation can be formally solved in terms of Jacobi elliptic functions (see for instance \cite{whittaker_watson}). Define the new variables $z = 2K(m)\, u$, $\omega = 2K(m)\, v$ and $z_i = 2K(m)\, \Delta_i$, where $K(m)$ is the complete elliptic integral of the first kind, and $m$ is the parameter (see Appendix~\ref{app: functions}). Then the curve can be rewritten as
\be
\label{eq:continuum}
\cn(z) \cn(\omega)  \cn(z-\omega) =  \cn(z_1) \cn(z_2) \cn(z_1 + z_2) \;.
\ee
We use the Jacobi elliptic functions $\cn(z,m)$, $\sn(z,m)$, $\dn(z,m)$ and keep the dependence on the parameter $m$ implicit. Then, employing addition formulas, we can isolate the dependence on $z$:
\be
\sn^2 \left( z - \frac\omega2 \right) = \frac{ \cn(z_1) \cn(z_2) \cn(z_1 + z_2) - \cn^2(\omega/2) \cn(\omega) }{ m \sn^2(\omega/2) \cn(z_1) \cn(z_2) \cn( z_1 + z_2) - \dn^2(\omega/2) \cn(\omega) } \;.
\ee
The two values for $u$ can finally be written in terms of the inverse function $\arcsn$:
\be
u_\pm(v) = \frac{v}{2} \pm \frac{1}{2K(m)} \arcsn \left( \sqrt{ \frac{ \cn(z_1) \cn(z_2) \cn(z_1 + z_2) - \cn^2(\omega/2) \cn(\omega) }{ m \sn^2(\omega/2) \cn(z_1) \cn(z_2) \cn(z_1 + z_2) - \dn^2(\omega/2) \cn(\omega) }} \, \right) \;.
\ee

The two equations $u = u_\pm(v)$ describe the curve as a double cover of the torus. The two sheets of the covering meet at all points where $u_+(v) = u_-(v) = v/2$. We could ask whether these are smooth points where the covering is branched, or rather are self-intersection points of the curve or intersection points of two connected components the curve is composed of. We can see that these must be smooth branch points with a simple argument. The points lie on the hyperplane $\theta_1(2u-v)=0$ on which $Q_2=1$. However, we could have equivalently described the curve as $v = v_\pm(u)$. If those points were intersection points, and so if they had an intrinsic geometric characterization, they would also solve $v_+(u) = v_-(u) = u/2$ and so would also lie on the hyperplane $\theta_1(u - 2v) = 0$ on which $Q_3 = 1$. However the intersections of the two hyperplanes (which intersect also the hyperplane $\theta_1(u+v)=0$) are precisely the HL solutions, and one can easily check that they do not lie on the curve (\ref{SU(3) curve}). This contradiction implies that there are no special intersection points on the curve, only smooth branch points. In particular this implies that the curve is composed of a single connected component.

More directly, we can compute the gradient of $h$. Exploiting the relations (\ref{more theta identities}) and (\ref{theta derivatives}), with some algebra we obtain that on the curve $h(u,v; \Delta,\tau) = 0$ it holds
\be
\partial_u h = - \frac{\theta_1'(0) \, \theta_1(2u-v)}{ \theta_1(u) \, \theta_1(u-v)} \, h(0,v; \Delta, \tau) \;,\qquad \partial_v h = \frac{\theta_1'(0) \, \theta_1(2v-u)}{ \theta_1(u) \, \theta_1(u-v)} \, h(u,0; \Delta, \tau) \;.
\ee
There are no points where $h = \partial_u h = \partial_v h = 0$, therefore all points on the curve are smooth.

\section*{Acknowledgements}

We thank Ofer Aharony, Arash Arabi Ardehali and Jan Troost for useful discussions and correspondence, and in particular Paolo Milan for collaboration in the early stage of this work. We also thank the organizers and participants of the SCGP seminar series on ``Supersymmetric black holes, holography and microstate counting'', where part of these results were announced, for interesting discussions.
F.B. is partially supported by the ERC-COG grant NP-QFT No.~864583, by the MIUR-SIR grant RBSI1471GJ, by the MIUR-PRIN contract 2015 MP2CX4, as well as by the INFN ``Iniziativa Specifica ST\&FI''.

\appendix

\section{Special functions}
\label{app: functions}

We use fugacities and chemical potentials related by
\be
z = e^{2\pi i u} \;,\qquad\qquad p = e^{2\pi i \sigma} \;,\qquad\qquad q = e^{2\pi i \tau} \;,
\ee
with $|p|, |q|<1$. The $q$-Pochhammer symbol is defined as
\be
(z;q)_\infty = \prod_{j=0}^\infty (1-z q^j) \;.
\ee
The elliptic theta function $\theta_0$ is defined as
\be
\theta_0(u;\tau) = (z;q)_\infty \, (q/z;q)_\infty \;.
\ee
The elliptic gamma function is defined as
\be
\wt\Gamma(u; \sigma, \tau) = \prod_{m,n=0}^\infty \frac{ 1 - p^{m+1} q^{n+1} / z}{ 1- p^m q^n z} \;.
\ee

We use the following definitions for the Jacobi theta functions:
\begin{align}
\theta_1(u;\tau) &= 2 \, e^{\pi i \tau/4} \sin(\pi u) \prod_{n=1}^\infty (1-q^n)(1-q^n z)(1-q^n/z) = i \, e^{\pi i \tau/4 -\pi i u} \, (q;q)_\infty \, \theta_0(u;\tau) \nn \\
\theta_2(u;\tau) &= 2 \, e^{\pi i \tau/4} \cos(\pi u) \prod_{n=1}^\infty (1-q^n)(1+q^nz)(1+q^n/z) = e^{\pi i \tau/4 -\pi i u } \, (q;q)_\infty \, \theta_0\bigl( u + \tfrac12;\tau \bigr) \nn \\
\theta_3(u;\tau) &= \prod_{n=1}^\infty (1-q^n) \bigl( 1 + q^{n-\frac12} z \bigr)\bigl( 1 + q^{n- \frac12} / z \bigr) = (q;q)_\infty \, \theta_0\bigl( u + \tfrac12 + \tfrac\tau2; \tau \bigr) \nn \\
\theta_4(u;\tau) &= \prod_{n=1}^\infty (1-q^n) \bigl( 1 - q^{n -\frac12} z \bigr) \bigl( 1 - q^{n - \frac12} / z \bigr) = (q;q)_\infty \, \theta_0 \bigl( u + \tfrac\tau2; \tau) \;.
\end{align}
Notice that $\theta_1$ is odd while $\theta_{2,3,4}$ are even under $u \to -u$.
One set of identities we use is
\bea
\label{more theta identities}
\theta_{1}(u+v) \, \theta_{2}(u-v) \, \theta_{3}(0) \, \theta_{4}(0) &= \theta_{1}(u) \, \theta_{2}(u) \, \theta_{3}(v) \, \theta_{4}(v) + \theta_{3}(u) \, \theta_{4}(u) \theta_{1}(v) \, \theta_{2}(v) \\
\theta_{1}(u+v) \, \theta_{3}(u-v) \, \theta_{2}(0) \, \theta_{4}(0) &= \theta_{1}(u) \, \theta_{3}(u) \, \theta_{2}(v) \, \theta_{4}(v) + \theta_{2}(u) \, \theta_{4}(u) \theta_{1}(v) \, \theta_{3}(v) \\
\theta_{1}(u+v) \, \theta_{4}(u-v) \, \theta_{2}(0) \, \theta_{3}(0) &= \theta_{1}(u) \, \theta_{4}(u) \, \theta_{2}(v) \, \theta_{3}(v) + \theta_{2}(u) \, \theta_{3}(u) \theta_{1}(v) \, \theta_{4}(v) \\
\theta_{2}(u+v) \, \theta_{3}(u-v) \, \theta_{2}(0) \, \theta_{3}(0) &= \theta_{2}(u) \, \theta_{3}(u) \, \theta_{2}(v) \, \theta_{3}(v) - \theta_{1}(u) \, \theta_{4}(u) \theta_{1}(v) \, \theta_{4}(v) \\
\theta_{3}(u+v) \, \theta_{4}(u-v) \, \theta_{3}(0) \, \theta_{4}(0) &= \theta_{3}(u) \, \theta_{4}(u) \, \theta_{3}(v) \, \theta_{4}(v) - \theta_{1}(u) \, \theta_{2}(u) \theta_{1}(v) \, \theta_{2}(v) \;,
\eea
see for instance \cite{Kharchev:2015}. Here $\theta_r(u) \equiv \theta_r(u; \tau)$. We also use
\be
\label{theta derivatives}
\partial_u \left( \frac{\theta_1(u)}{\theta_4(u)} \right) = \pi \, \frac{\theta_4(0)^2 \, \theta_2(u) \, \theta_3(u) }{ \theta_4(u)^2}
\ee
and a similar relation with $2$ and $4$ exchanged, as well as $\theta_1'(0) = \pi \, \theta_2(0) \, \theta_3(0) \, \theta_4(0)$.

The Jacobi elliptic functions are related to the Jacobi theta functions by
\bea
\cn\bigl( 2K(m)u ,\, m \bigr) &= \biggl( \frac{1-m}m \biggr)^{1/4} \, \frac{\theta_2(u; \tau)}{\theta_4(u;\tau)} \\
\sn\bigl( 2K(m)u ,\, m \bigr) &= \biggl( \frac1m \biggr)^{1/4} \; \frac{\theta_1(u; \tau)}{\theta_4(u;\tau)} \\
\dn\bigl( 2K(m)u ,\, m \bigr) &= (1-m)^{1/4} \, \frac{\theta_3(u; \tau)}{\theta_4(u;\tau)} \;.
\eea
The parameter $m$ is related to the modulus $\tau$ by
\be
m(\tau) = \frac{\theta_2(0;\tau)^4 }{ \theta_3(0;\tau)^4} \;.
\ee
The function $K(m)$ is the complete elliptic integral of the first kind, and is related to $\tau$ by
\be
\tau = i \, \frac{K(1-m)}{K(m)} \qquad\text{or}\qquad 2K(m) = \frac{\theta_3(0; \tau)^2}\pi \;.
\ee
We use the addition formula
\be
\cn(x+y,m) \cn(x-y,m) = \frac{ \cn^2(y,m) - \sn^2(x,m) \dn^2(y,m) }{ 1 - m \sn^2(x,m) \sn^2(y,m) } \;.
\ee


\bibliographystyle{ytphys}
\bibliography{BHEntropy}


\end{document}